\documentclass[aps,english,superscriptaddress,preprintnumbers,reprint,nofootinbib,amsmath,amssymb,prl]{revtex4-1}
\usepackage[version=3]{mhchem}
\usepackage{natbib}
\usepackage{graphicx}% Include figure files
\usepackage{dcolumn}% Align table columns on decimal point
\usepackage{bm}% bold math
\usepackage{hyperref}% add hypertext capabilities
\hypersetup{
    colorlinks=true,
    linkcolor=blue,
    filecolor=blue,
    urlcolor=blue,
   citecolor=blue,
}

\usepackage{enumitem}
\begin{document}

%\preprint{APS/123-QED}

\title{Interaction induced hybridization of Majorana zero-modes in a coupled quantum-dot hybrid-nanowire system}% Force line breaks with \\

\author{L. S. Ricco}
\email[corresponding author:]{luciano.silianoricco@gmail.com}
\affiliation{S\~ao Paulo State University (Unesp), School of Engineering, Department of Physics and Chemistry, 15385-000, Ilha Solteira-SP, Brazil}

\author{Y. Marques}
\affiliation{ITMO University, St.~Petersburg 197101, Russia}

\author{J. E. Sanches}
\affiliation{S\~ao Paulo State University (Unesp), School of Engineering, Department of Physics and Chemistry, 15385-000, Ilha Solteira-SP, Brazil}

\author{I. A. Shelykh}
\affiliation{ITMO University, St.~Petersburg 197101, Russia}
\affiliation{Science Institute, University of Iceland, Dunhagi-3, IS-107,
Reykjavik, Iceland}

\author{A. C. Seridonio}
\email[corresponding author:]{acfseridonio@gmail.com}
\affiliation{S\~ao Paulo State University (Unesp), School of Engineering, Department of Physics and Chemistry, 15385-000, Ilha Solteira-SP, Brazil}
\affiliation{S\~ao Paulo State University (Unesp), IGCE, Department of Physics, 13506-970, Rio Claro-SP, Brazil}

\date{\today}% It is always \today, today,

\begin{abstract} 
We study the low-energy transport properties of a hybrid device composed by a native quantum dot coupled to both ends of a topological superconducting nanowire section hosting Majorana zero-modes. The account of the coupling between the dot and the farthest Majorana zero-mode allows to introduce the topological quality factor, characterizing the level of topological protection in the system. We demonstrate that Coulomb interaction between the dot and the topological superconducting section leads to the onset of the additional overlap of the wavefunctions describing the Majorana zero-modes, leading to the formation of trivial Andreev bound states even for spatially well-separated Majoranas. This leads to the spoiling of the quality factor and introduces a constraint for the braiding process required to perform topological quantum computing operations.
\end{abstract}


\maketitle

%\tableofcontents

\textit{Introduction.---} Errors coming from quantum decoherence are undoubtedly one of the most significant obstacles for successful realization of a reliable quantum computer. In this regard, topological quantum computing has been considered as an attractive solution for overcoming of the related problems ~\cite{Alicea2011,Kitaev2003,Freedman2003,Knapp2016}. It processes quantum information in a nonlocal fashion, and in addition exploits the peculiarities of non-Abelian braiding statistics~\cite{RevNonabelian2008}, allowing the performence of decoherence-free and fault-tolerant quantum logical operations~\cite{Aasen2016}. In this context, so-called Majorana zero-modes (MZMs) emerging at opposite ends of a 1D spinless \textit{p}-wave superconductor~\cite{Kitaev2001, RevMajoranaAlicea, RevMajoranaFranz,RevMajoranaAguado} exhibiting pronounced non-Abelian behaviour associated topological degeneracy~\cite{RevNonabelian2008,Clarke2017} have been proposed as elementary building blocks of a topological quantum computing hardware~\cite{Plugge2017}.

Although \textit{p}-wave superconductivity is very rare in nature~\cite{RevModPhysSR2RUO4,Moore1991,Kopnin1991,Volovik1999}, it can be engineered in quasi-1D semiconducting nanowires with strong Rashba spin-orbit coupling, brought in close proximity with conventional superconductor and placed in strong external magnetic field parallel to the spin-orbit intrinsic field~\cite{LutchynPRL2010,RevMajoranaAlicea,RevMajoranaFranz,RevMajoranaAguado}. In this configuration, MZMs manifest themselves at the opposite ends of topological superconducting segment~\cite{LutchynPRL2010,RevMajoranaAguado}. They can be probed by means of tunneling spectroscopy experiments~\cite{LutchynReviewMat2018}, where the appearance of a robust zero-bias anomaly (ZBP), verified in a set of  experiments~\cite{MourikScience2012,KrogstrupNatMater2015,AlbrechtNature2009,DengScience2016,Nichele2017,ZhangNatNanotech2018}, was considered as direct evidence of their presence.

\begin{figure}[t]
	\centerline{\includegraphics[width=3.0in,keepaspectratio]{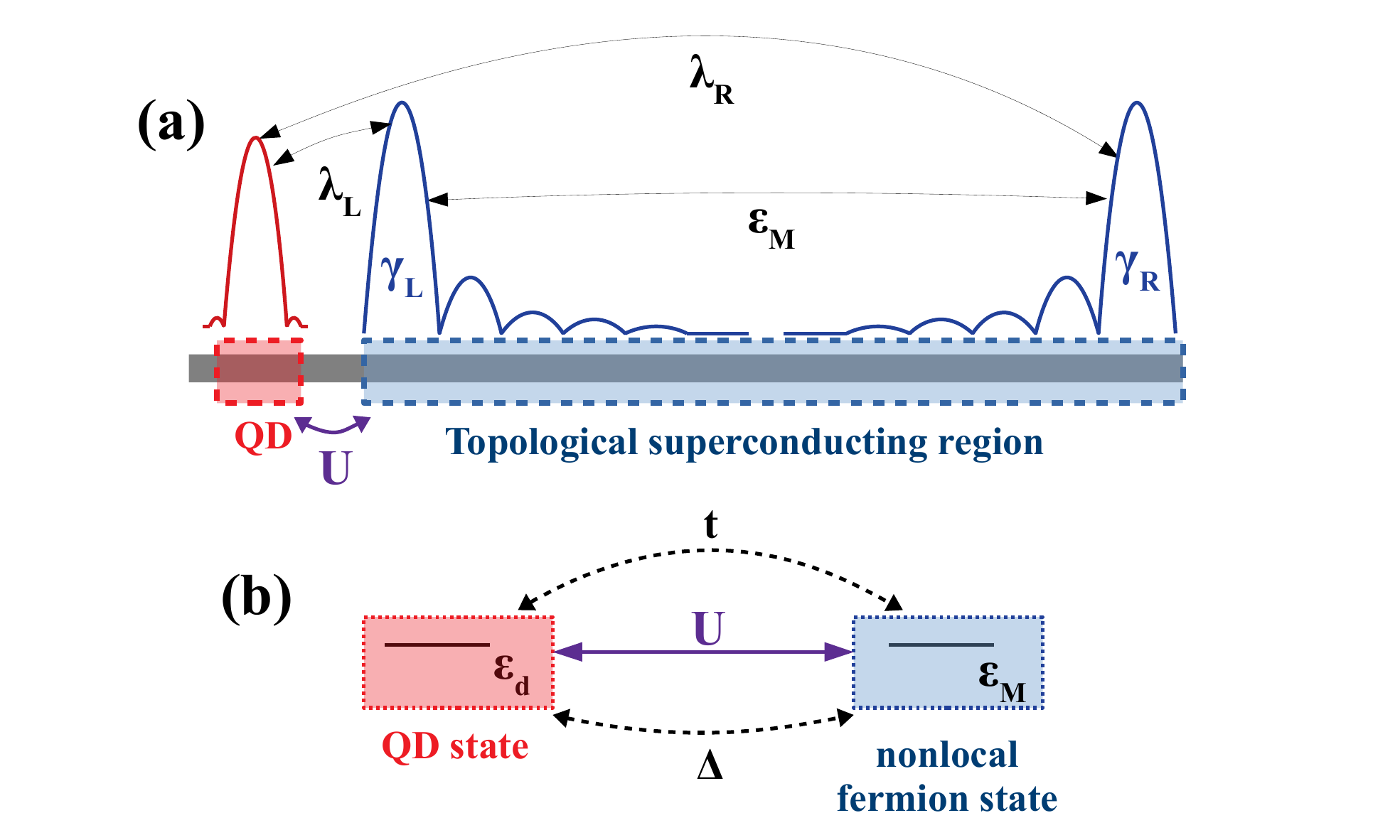}}\caption{\label{fig:Device}(a) The sketch of the considered device, consisting of a hybrid topological superconducting nanowire in spinless regime (blue region), hosting  left ($\gamma_{L}$) and right ($\gamma_{R}$) MZMs at the opposite ends, which are both coupled to a single level quantum dot (red region). $\lambda_{L}$ and $\lambda_{R}$ give the coupling amplitudes between the QD and left/right MZMs, respectively, while the parameter $\varepsilon_{M}$ characterizes the spatial overlap between the wave functions describing the MZMs. There is also a Coulomb repulsion $U$ between the QD and MZMs. (b) Equivalent scheme of the device. The topological superconducting section is now represented in terms of a nonlocal fermionic site with energy $\varepsilon_{M}$, constructed as a linear combination of spatially separated MZMs. In this formulation, $U$ gives the strength of the intersite Coulomb charging repulsion between the QD and the nonlocal fermion. Moreover, the QD connects to the fermionic site via normal tunneling and crossed Andreev reflection terms, with amplitudes $t$ and $\Delta$, respectively.}
\end{figure}

However, another physical mechanisms can be responsible for the formation of ZBPs, as \textit{e.g.}, Kondo effect~\cite{CronenwettScience1998,Goldhaber1998} and formation of Andreev bound states (ABSs)~\cite{KellsPRB2012,LeeNanotech,LiuPRB2017,LiuPRB2018,HellPRB2018,LaiPRB2019,ChenPRL2019,Marra2019,Stanescu2019,Vuik2019}. These latter can be viewed as overlapping MZMs which can remain pinned at zero-energy for a wide range of tunable parameters and even show perfect quantized conductance patterns~\cite{Pan2020,Haining2020}, mimicking exactly the behaviour expected for topologically protected MZMs. Strong disorder can also lead to such trivial ZBPs~\cite{BagretsPRL2012,Pan2020}. In order to distinguish between the ZBPs stemming from authentic topological MZMs and those arising from trivial zero-energy ABSs, a plethora of strategies have been proposed~\cite{Scuray2017,DJClarcke2017,Prada2017,DengPRB2018,Moore2018,LiuPRB2018,Sanches2020,Ricco2020topological,Yu2020nonmajorana}. Among them, we draw attention for those ones which suggest experimental verification of ``how topological'' are the MZMs, either by means of the measurement of so-called topological quality factor~\cite{DJClarcke2017} or degree of Majorana nonlocality.~\cite{Prada2017,DengPRB2018}.      

In the present paper, we analyze the effect of the Coulomb interaction between electrons located in QD and MZMs on the above mentioned quantities. We study the low-energy spectrum of a hybrid system composed by a quantum dot (QD) coupled to a topological superconducting section hosting MZMs at the opposite ends, as it is sketched in Fig.~\ref{fig:Device}(a). Our findings reveal that the Coulomb interaction between the QD and the superconducting section leads to the additional overlap of the wave functions describing the individual MZMs, even when they are spatially far apart. Hence, the information about either the topological quality factor or degree of Majorana nonlocality cannot be unambiguously extracted in the interacting case. We discuss the effects of the Coulomb charging energy on topological operations, demonstrating that a braiding step must be performed diabatically with respect to the Coulomb strength in order to be considered as being topologically protected.

\textit{Model and Methods.---} Consider a hybrid nanowire in the spinless regime~\cite{Kitaev2001} (\textit{i.e.,} when Zeeman splitting induced by strong external magnetic field parallel to the axis of the wire exceeds all other characteristic energies in the system),  composed by a native quantum dot~\cite{DengScience2016,DengPRB2018} with a single level coupled to both left and right MZMs located at the ends of the topological superconducting section of the corresponding nanowire~\cite{VernekPRB2014,DJClarcke2017,Prada2017,RiccoOscillations2018,RiccoSciReports}, see Fig.~\ref{fig:Device}(a). The effective low-energy Hamiltonian describing such a system reads
\begin{eqnarray}
    \mathcal{H} & = & \varepsilon_{d}c^{\dagger}_{d}c_{d} + \imath\varepsilon_{M}\gamma_{L}\gamma_{R}+\lambda_{L}(c_{d}-c^{\dagger}_{d})\gamma_{L}\nonumber\\
     & + & \lambda_{R}(c_{d}+c^{\dagger}_{d})\gamma_{R} + H_{U}, \label{eq:H_MZM_basis}
\end{eqnarray}
where the operator $c^{\dagger}_{d}(c_{d})$ creates (annihilates) an electron at the QD with energy $\varepsilon_{d}$ and $\gamma_{L}(\gamma_{R})$  describes the left (right) Majorana mode emerging at the ends of the topological superconducting section. The Majorana operators satisfy self-conjugation $\gamma_{i}=\gamma_{i}^{\dagger}$, with $\lbrace \gamma_{i},\gamma_{j} \rbrace=\delta_{i,j}(i=L,R)$~\cite{RevMajoranaAguado}. The amplitude of the spatial overlap between these MZMs is given by $\varepsilon_{M}$. 

The last term in the Hamiltonian corresponds to the Coulomb repulsion between the electron of the QD, and MZMs. It can be easily constructed using the representation of a nonlocal fermion, which is a linear combination of two MZMs, $c^{\dagger}_{f}=(\gamma_{L}-\imath\gamma_{R})/\sqrt{2}$, $n_{f}=c^{\dagger}_{f}c_{f}$. In this picture, the interaction can be simply reads as $H_{U}=Un_{d}n_{f}$, where $n_d=c_d^\dagger c_d$ is an operator of the occupancy of the QD. Note that expressing the Majorana operators in terms of the fermionic ones, \textit{i.e.,} using the transformation $\gamma_{L}=(c^{\dagger}_{f}+c_{f})/\sqrt{2}$ and $\gamma_{R}=\imath(c^{\dagger}_{f}-c_{f})/\sqrt{2}$, the interaction term reads:
\begin{equation}
 H_{U}=Un_{d}n_{f}=Un_d\left(i\gamma_L\gamma_R+\frac{1}{2}\right).
 \label{HU}
 \end{equation}
It can be easily noted that the structure of this term closely resembles those for the second term in the Eq.~(\ref{eq:H_MZM_basis}), $\imath\varepsilon_{M}\gamma_{L}\gamma_{R}$, which describes the hybridization of MZMs, leading to the formation of topologically trivial ABSs. Therefore, the QD-MZM repulsion leads to additional hybridization of the MZMs, whose amplitude $Un_d$ depends on the QD occupancy. This has a dramatic effect on topological characteristics of the system, as we will show below.
Note that the intradot Coulomb interaction can be safely neglected, since only single occupancy is allowed in the QD within the spinless regime. 

The system Hamiltonian [Eq.~(\ref{eq:H_MZM_basis})] can be rewritten in terms of nonlocal fermionic operators $c^{\dagger}_{f}(c_{f})$ as~\cite{Leijnse2012,RevMajoranaAguado,RiccoSciReports}
\begin{eqnarray}
    \mathcal{H} & = & \varepsilon_{d}c^{\dagger}_{d}c_{d} + \varepsilon_{M}c^{\dagger}_{f}c_{f}+ (tc_{d}c^{\dagger}_{f} + \Delta c_{d}c_{f} + \text{H.c.}) \nonumber \\
     & + & H_{U} + \text{const.},\label{eq:H_fermionic_basis}
\end{eqnarray}
where $\lambda_{L}=(t+\Delta)/\sqrt{2}$, $\lambda_{R}=\imath(\Delta-t)/\sqrt{2}$ and $\varepsilon_{M}$ has now the meaning of a nonlocal fermionic excitation energy. The transport between the QD and topological wire is described in terms of a normal tunneling and crossed Andreev reflection with $t$ and $\Delta$ being the corresponding amplitudes, which are both chosen real~\cite{RiccoSciReports,Leijnse2012,DJClarcke2017}. The MZMs are quadratically protected in the ``sweet spot'' $t=\Delta$ for $\varepsilon_{M}=U=0$~\cite{Leijnse2012}, which corresponds to the ideal situation of well-isolated MZMs.

The system Hamiltonian given by Eq.~(\ref{eq:H_fermionic_basis}) can be recast in the matrix form as:
\begin{equation}
\mathcal{H}=\begin{pmatrix}0 & 0 & 0 & \Delta\\
0 & \varepsilon_{M} & t & 0\\
0 & t & \varepsilon_{d} & 0\\
\Delta & 0 & 0 & \varepsilon_{d}+\varepsilon_{M}+U \label{eq:H_matrix}
\end{pmatrix}
\end{equation}
where we have chosen the following basis $\lbrace |00\rangle, |01\rangle, |10\rangle, |11\rangle \rbrace$ for the number states $|n_{d}n_{f}\rangle$~\cite{DJClarcke2017,Leijnse2012}. 

The conductance through the system is determined by the density of states of the QD, which in the Lehmann representation reads~\cite{Bruus}:
\begin{equation}
\rho_{d}(V) = \frac{1}{\mathcal{Z}}\sum_{n,m}|\langle m |c_{d}^{\dagger}| n \rangle|^{2}(e^{\beta E_{n}} + e^{\beta E_{m}})\delta(V + E_{n} - E_{m}),\label{eq:Lehmann}
\end{equation} 
where $\beta=1/k_{B}T$, $\mathcal{Z}$ is the partition function of Eq.~(\ref{eq:H_matrix}) with a complete set of eingenstates $\lbrace  \left| (m)n \right\rangle \rbrace$ and associated eigenvalues $E_{m(n)}$. For $T=0$, the differential conductance through the QD $dI/dV\propto\rho_{d}(V)$ can be probed with a pair of metallic leads, with $V$ being the corresponding bias-voltage. Note that the attachment of metallic leads will broaden the peaks in the density of states, and the delta functions in Eq.~(\ref{eq:Lehmann}) should be replaced by lorentzians. In our further analysis, however, we will be focusing on the position of the conductance peaks only and this broadening will be neglected. 

Eq.~(\ref{eq:Lehmann}) determines that a peak is registered in the conductance through the QD whenever a transition between the ground state of the system [Eq.~(\ref{eq:H_matrix})] and an excited state with opposite parity is allowed~\cite{DJClarcke2017}, \textit{i.e.,} when the matrix elements $|\langle m |c_{d}^{\dagger}| n \rangle|\neq 0$, and for $V = (E_{m}-E_{n})$.  

By taking into account the four possible eigenvalues of the QD-topological superconducting section extracted from Eq.~(\ref{eq:H_matrix}) and considering the transition rules defined by Eq.~(\ref{eq:Lehmann}), it is straightforward to show that a conductance peak will emerge at the low-energy spectrum of the system for $V=V_{1}+V_{2}$, with 
\begin{equation}
V_{1}  =  \pm\frac{1}{2}[\sqrt{\epsilon_{-}^{2}+(2t)^{2}}+\sqrt{(\epsilon_{+}+U)^{2}+(2\Delta)^{2}}- U] \label{eq:V1}
\end{equation}
and
\begin{equation}
V_{2} = \pm  \frac{1}{2}[\sqrt{\epsilon_{-}^{2}+(2t)^{2}}-\sqrt{(\epsilon_{+}+U)^{2}+(2\Delta)^{2}}- U], \label{eq:V2}    
\end{equation}
where $V_{1}$ describes the energy spectrum corresponding to the QD states (dashed green lines), while $V_{2}$ corresponds to the MZMs spectrum (red lines), with $\epsilon_{\pm}=\varepsilon_{d}\pm \varepsilon_{M}$. For $U=0$, we just recover the case described by Clarke~\cite{DJClarcke2017}, who also used the Lehmann representation for the noninteracting case and whose work is used as the reference point in our study.

Before presenting our results, it is worth to  discuss shortly the so-called topological quality factor, defined as:
\begin{equation}
\mathcal{Q}=1-\frac{|\lambda_{R}|}{|\lambda_{L}|},
\end{equation}
proposed by Clarke~\cite{DJClarcke2017} as a quantitative criterium allowing to estimate if a system can be interpreted as topological for $\varepsilon_{M}\approx 0$. Well-separated MZMs corresponds to the ideal topologically protected situation, wherein $\lambda_{R}=\varepsilon_{M}=0$, yielding the highest quality factor $\mathcal{Q}=1$. As the nonlocal nature of MZMs fades away, $\lambda_{R}$ enhances and $\mathcal{Q}$ approaches zero, the crossover to topologically trivial case of strongly overlapping MZMs, \textit{i.e.,} ABSs, occurs. The quality factor is related with the so-called degree of Majorana nonlocality introduced by Prada \textit{et al.}~\cite{Prada2017} and experimentally studied by Deng \textit{et al.}~\cite{DengPRB2018}. 

\textit{Noninteracting case.---} To understand better the role of the Coulomb repulsion between the electrons from QD and MZMs, we start from a brief discussion of the results for the noninteracting case ($U=0$). 

\begin{figure}[t]
	\centerline{\includegraphics[width=3.3in,keepaspectratio]{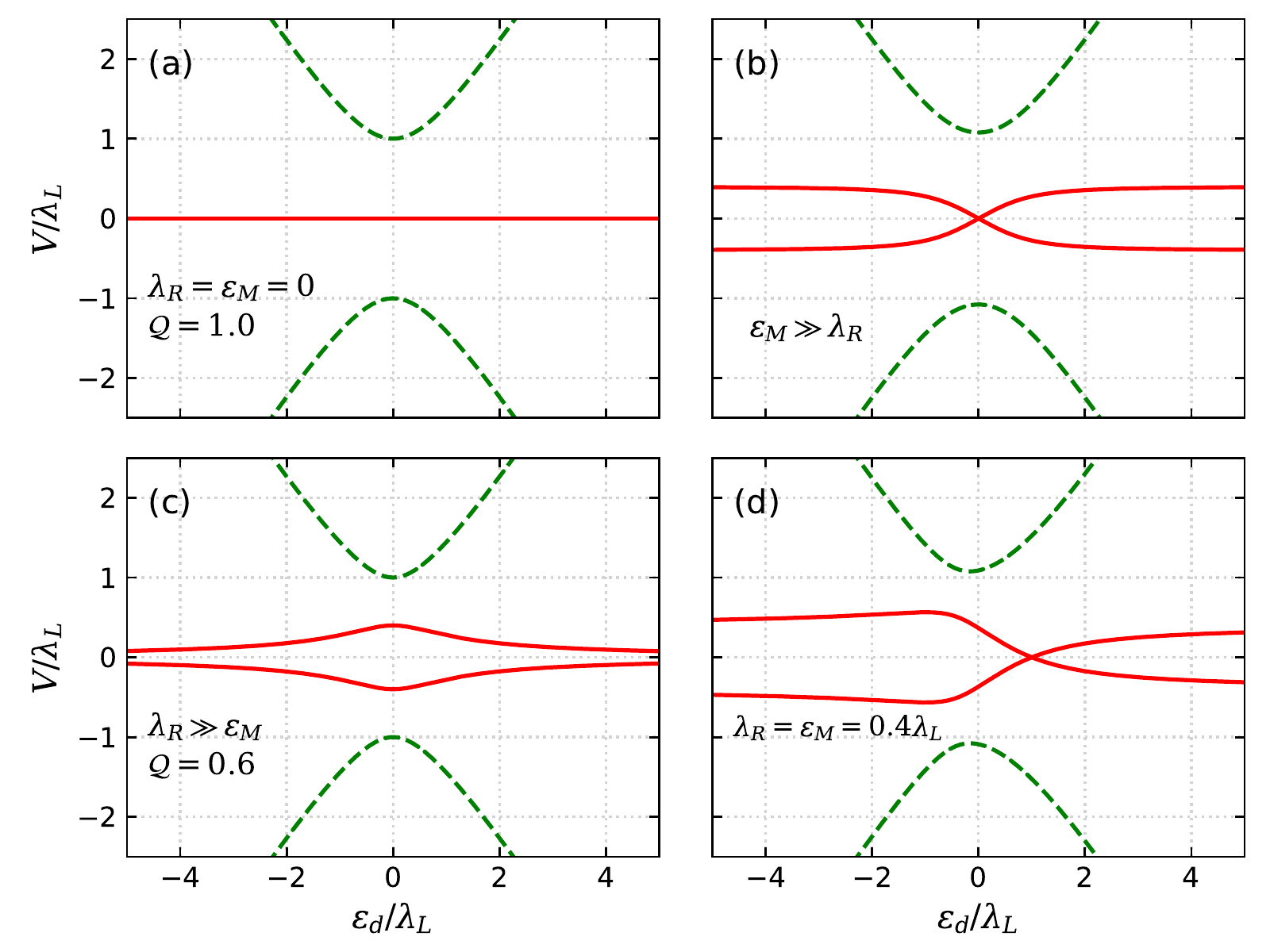}}
	\caption{\label{fig:Result1}Low-energy spectrum characterizing the conductance profiles [Eqs.~(\ref{eq:V1}) and~(\ref{eq:V2})] as a function of the QD energy-level $\varepsilon_{d}$ and applied bias-voltage $V$ for non interacting case ($U = 0$). The red lines describe the position of the peak, related to the states of the topological superconducting section hosting MZMs,  dashed-green lines correspond to the QD states. (a) The case of well-separated MZMs ($\lambda_{R}=\varepsilon_{M}=0$), yielding the highest topological quality factor, $\mathcal{Q}=1$. Horizontal red line corresponds to the robust ZBP. (b) The case, when the wavefunctions describing the MZMs at the opposite ends of the topological section are strongly overlapping ($\varepsilon_{M}\gg \lambda_{R}$). ZBP is splitted, producing characteristic bowtie, or double-fork, pattern (c) The case of a finite overlap between the QD and the right MZM, with $\mathcal{Q}=0.6$, $\varepsilon_M=0$. (d) The case, when the overlap between the MZMs is comparable with those between the QD and the right MZM.}
\end{figure}

In Fig.~\ref{fig:Result1}, the corresponding low-energy spectrum is presented. The ideal situation of well-separated MZMs is shown in Fig.~\ref{fig:Result1}(a). There is no overlap neither between the wave function of the QD and that one describing the right MZM ($\lambda_{R}=0$), nor between the wave functions describing the right and left MZMs ($\varepsilon_{M}=0$), and $\mathcal{Q}=1$. This corresponds to the perfect topological case with quadratically protected MZMs~\cite{Leijnse2012}, for which there is a robust single zero-bias conductance peak which remains pinned to zero when the QD energy-level is tuned (red horizontal line). 

The situation drastically changes when a finite overlap between the MZMs is taken into account ($\varepsilon_{M}\neq0$), as it is shown in Fig.~\ref{fig:Result1}(b). In this case, the spectrum corresponding to the MZMs (red lines) splits, revealing the  crossing at $\varepsilon_{d}=0$, the pattern known as \textit{bowtie}~\cite{Prada2017} or \textit{double-fork}  profile \cite{RiccoSciReports}.

Fig.~\ref{fig:Result1}(c) illustrates the case when the overlap amplitude $\lambda_{R}$ between the QD and the right MZM is finite and $\varepsilon_{M}\approx 0$, and the corresponding topological quality factor is $\mathcal{Q}=0.6$. Te fact that $\mathcal{Q}<1$ introduces a constraint for topological quantum operations, as we shall see later. This quasi-MZMs are characterized by a \textit{diamond-like} profile~\cite{Prada2017,DengPRB2018} for the splitted ZBP. 

Finally, in Fig.~\ref{fig:Result1}(d) we illustrate the case where $\lambda_{R}=\varepsilon_{M}$ and highly asymmetric \textit{bowtie} is revealed. Both Figs.~\ref{fig:Result1}(c) and (d) correspond quite well to the experimental conductance profiles reported by Deng.\textit{et al.}~\cite{DengScience2016}, which means that the experimental prototype was in fact in the trivial phase~\cite{DJClarcke2017,Prada2017}.

\textit{Interacting case.---} The low-energy spectrum of the system drastically changes when the Coulomb interaction $U$ between the QD and the topological superconducting section hosting MZMs is taken into account. 

\begin{figure}[t]
	\centerline{\includegraphics[width=3.3in,keepaspectratio]{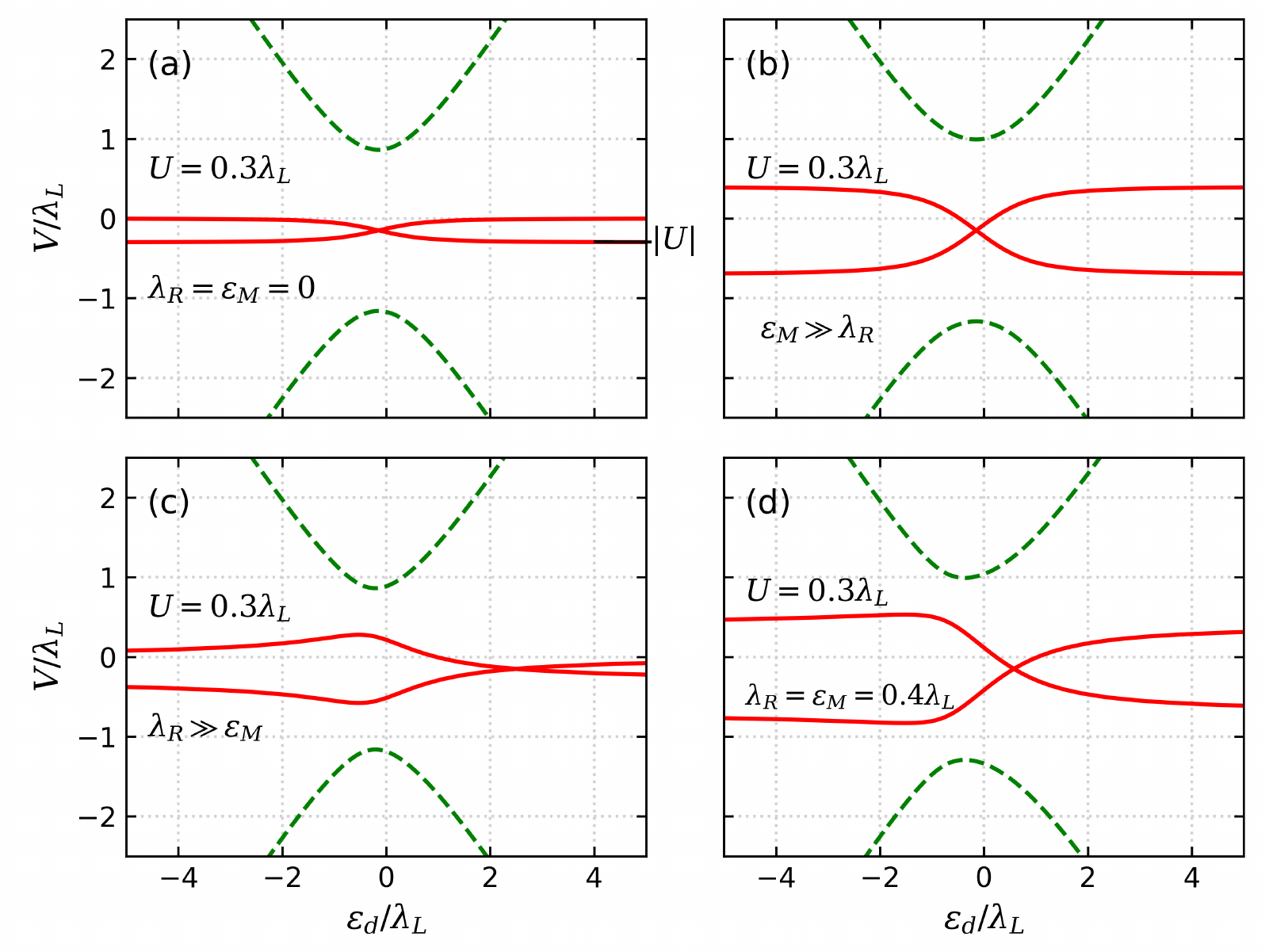}}
	\caption{\label{fig:Result2} Low-energy spectrum characterizing the conductance profiles [Eqs.~(\ref{eq:V1}) and~(\ref{eq:V2})] as a function of the QD energy-level $\varepsilon_{d}$ and applied bias-voltage $V$ for the interacting case ($U\neq0$). The red lines describe the position of the peak, related to the states of the topological superconducting section hosting MZMs,  dashed-green lines correspond to the QD states. (a) The case of well-separated MZMs ($\lambda_{R}=\varepsilon_{M}=0$), yielding the highest topological quality factor, $\mathcal{Q}=1$. The robust ZBA, represented by horizontal red line in Fig.~\ref{fig:Result1}(a) is destroyed by the intersite repulsion, and transforms into bowtie profile. (b) The case, when the wavefunctions describing the MZMs at the opposite ends of the topological section are strongly overlapping ($\varepsilon_{M}\gg \lambda_{R}$). The finite $U$ leads the appearance of slight asymmetry in the \textit{bowtie} profile, and greater splitting of the lines. (c) The case of a finite overlap between the QD and the right MZM, $\lambda_R\neq0,\varepsilon_M=0$. The symmetric diamond profile, characteristic to the non interacting case, transforms into asymmetric bowtie due to interaction induced hybridization of MZMs. (d) The case, when the overlap between the MZMs is comparable with those between the QD and the right MZM. Interactions does not change qualitatively the figures, increasing the degree of asymmetry and splitting.}
\end{figure}

As it was already mentioned, in the representation of nonlocal fermions this interaction describes a simple charge repulsion, while in the basis of MZMs it has the structure similar to those of the term describing the overlap between the MZMs, see Eq.~(\ref{HU}). This leads to the splitting of the ZBP and formation of a bowtie structure even for the seemingly ideal case, corresponding to $\lambda_R=\varepsilon_M=0$, as it is shown in Fig.~\ref{fig:Result2}(a). This unexpected loss of robustness of the ZBP is precisely owing to the Coulomb correlation, which effectively overlaps the wavefunctions of the MZMs having the QD as an intermediate, in accordance with Eq.~(\ref{HU}). Note that as in the considered case the amplitude of the overlap depends on the occupancy of the dot, the corresponding \textit{bowtie} structure becomes slightly distorted, forming a crooked \textit{double-fork} with crossing point slightly shifted from $\varepsilon_d=0$. The information about the strength of Coulomb interaction can be extracted from such a low-energy spectrum profile, as indicated in the corresponding Fig.~\ref{fig:Result2}(a). 

If the overlap between the MZMs is already present ($\varepsilon_M\neq0$), Coulomb correlations contribute to the additional vertical distancing of the \textit{bowtie} lines, together with a horizontal shift of the crossing point from $\varepsilon_d=0$, as it is shown in Fig.~\ref{fig:Result1}(b).

Fig.~\ref{fig:Result2}(c) illustrates the regime for which the concept of the quality factor was originally introduced, namely $\lambda_R\gg\varepsilon_M$. One can clearly see that the account of the Coulomb corrections destroys the symmetric \textit{diamond} shape characteristic to the noninteracting case and leads to the appearance of the crossing and formation of an asymmetric \textit{bowtie} profile, which is a direct consequence of the interaction induced hybridization of MZMs. As for the case when $\lambda_R$ and $\varepsilon_M$ are comparable, the situation is qualitatively similar for noninteracting and interacting cases, the quantitative difference being that the asymmetry of the profiles and the splitting between the lines is enhanced in the interacting regime, see Fig.~\ref{fig:Result2}(d). 

\begin{figure}[t]
	\centerline{\includegraphics[width=3.6in,keepaspectratio]{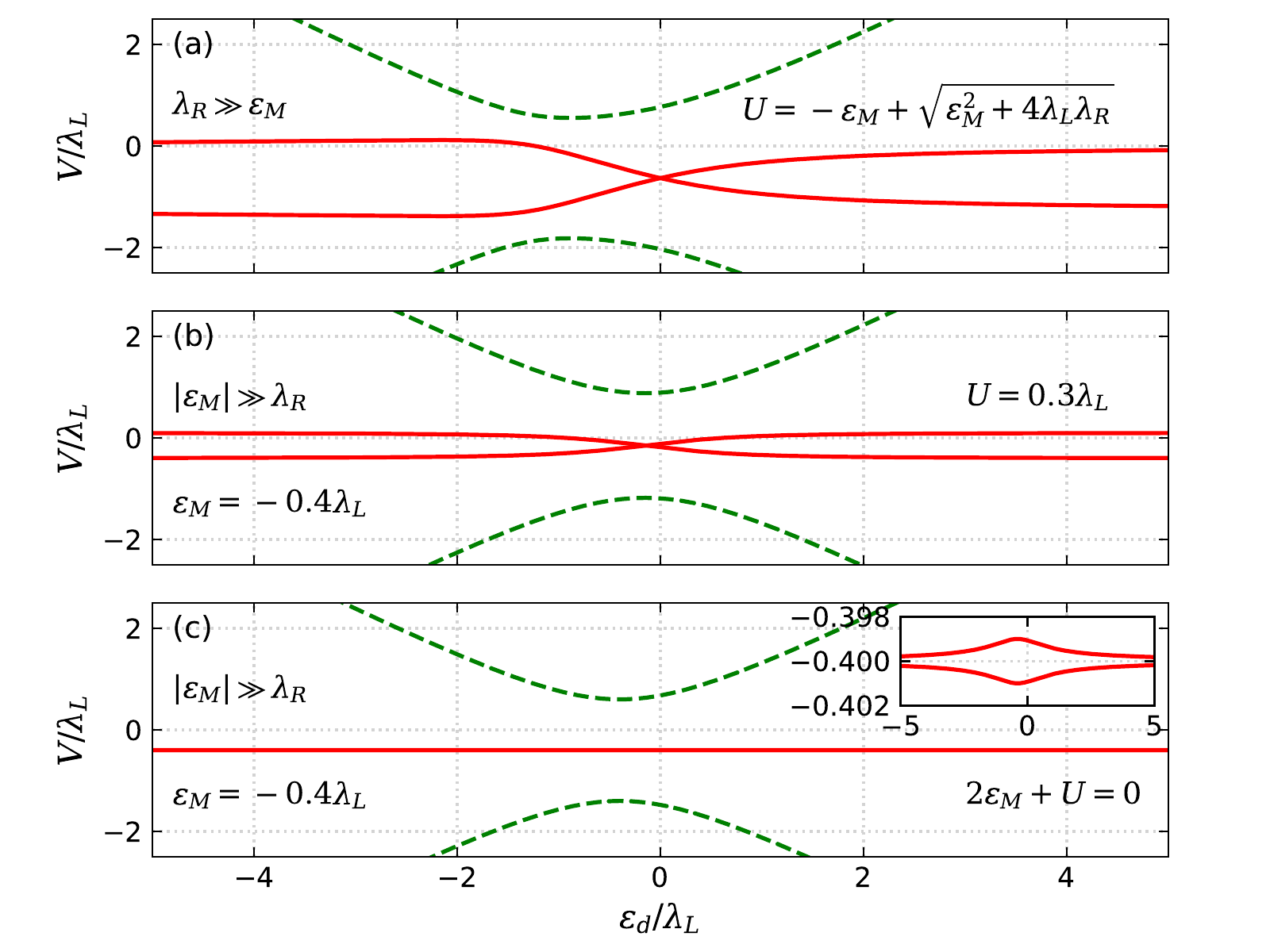}}
	\caption{\label{fig:Result3} Low-energy spectrum characterizing the conductance profiles [Eqs.~(\ref{eq:V1}) and~(\ref{eq:V2})] as a function of the QD energy-level $\varepsilon_{d}$ and applied bias-voltage $V$ for the interacting case ($U\neq0$). The red lines describe the position of the peak, related to the states of the topological superconducting section hosting MZMs, dashed-green lines correspond to the QD states. (a) The case wherein the value of $U$ corresponds to $\varepsilon_{\text{cross}}=0$ [Eq.~(\ref{eq:crossing_point})] for $\varepsilon_{M}$ and $\lambda_{R}\neq0$, showing a symmetric profile in relation to $\varepsilon_{d}=0$. (b) The overlap strength between the MZMs is negative, but holds the condition $|\varepsilon_{M}| \gg \lambda_{R}$. (c) Same situation of the panel above, but for the condition $2\varepsilon_{M} + U =0$, yielding an indistinguishable splitting of the states describing the MZMs. The zoomed spectrum near $V=0$ is shown in the inset panel.}
\end{figure}

Using Eqs.~(\ref{eq:V1}) and~(\ref{eq:V2}), one can obtain the expression for the position of the crossing of the red lines in Figs.~\ref{fig:Result1} and \ref{fig:Result2}:
\begin{equation}
\varepsilon_{\text{cross}} = \frac{2\lambda_{L}\lambda_{R}}{(2\varepsilon_{M} + U)} - \frac{U}{2}. \label{eq:crossing_point}
\end{equation}
For $\varepsilon_{M}\rightarrow 0, U \rightarrow 0$ and $\lambda_{L}\neq 0$, $\varepsilon_{\text{cross}}\rightarrow \infty$ and then, there is no crossing between the near-zero-energy states, as indeed one can verify in Fig.~\ref{fig:Result1}(c). However, for the interacting case wherein $U=0.3\lambda_{L}$, $\lambda_{R}=0.4\lambda_{L}$ and $\varepsilon_{M} \approx 0$, we obtain $\varepsilon_{\text{cross}} \approx 2.5\lambda_{L}$, which corresponds to Fig.~\ref{fig:Result2}(c).

Eq.~(\ref{eq:crossing_point}), one can verify that $U = -\varepsilon_{M} + \sqrt{\varepsilon_{M}^{2} + 4\lambda_{L}\lambda_{R}}$ gives the physical solution for $\varepsilon_{\text{cross}}=0$, with $\varepsilon_{M}$ and $\lambda_{L}\neq 0$, as shown in Fig.~\ref{fig:Result3}(a). By fulfilling this condition, the asymmetric profile of the near-zero-energy states in relation to $\varepsilon_{d}=0$ [Fig.~\ref{fig:Result2}(c)] is symmetrized.

Fig.~\ref{fig:Result3}(b) depicts the case of $\varepsilon_{M}<0$ and $|\varepsilon_{M}|\gg \lambda_{R}$. In the Majorana basis [Eq.~(\ref{eq:H_MZM_basis})], this situation corresponds to a negative overlap strength between the MZMs, once $\varepsilon_{M}$ is proportional to a cosine function, showing an oscillatory behaviour for shorter nanowires~\cite{RiccoOscillations2018}. Comparison between Figs.~\ref{fig:Result2}(b) and~\ref{fig:Result3}(b) shows that the negative value of $\varepsilon_{M}$ reduces the width of the near-zero-energy splitting in the interacting picture. This feature points out the the splitting strength of the states which corresponds to the MZMs (red lines) is proportional to $\varepsilon_{M} + U$. If one consider \textit{e.g.} the condition $2\varepsilon_{M} + U=0$ with $\varepsilon_{M}<0$ [Fig.~\ref{fig:Result3}(c)], the splitting is indistinguishable (see the zoomed inset panel), yielding an apparent plateau shifted from $V=0$.

\textit{Effects of the interaction on topological operations.---} Both the quality factor $\mathcal{Q} < 1$ ($\lambda_{R} \neq 0$) and spatial overlap $\varepsilon_{M}\neq 0$ between the MZMs introduce some restrictions for performing topological operations given by braiding processes~\cite{Alicea2011,Sau2011,Halperin2012,Aasen2016,DJClarcke2017}. Specifically, an exchange operation involving MZMs (braid step) must be performed very quickly with respect to characteristic times defined by the coupling $\lambda_{R}$ and the overlap $\varepsilon_{M}$, so that the system responds as if it were topologically protected ($\lambda_{R}=\varepsilon_{M}=0$, $\mathcal{Q}=1$). Once Coulomb correlations in the considered system induce the additional overlap of the MZMs [see Eq.~(\ref{HU})], it is natural to infer that a braiding operation also must be performed diabatically in relation to $U$. All these constraints imply that~\cite{DJClarcke2017}
\begin{equation}
\frac{\hbar}{\lambda_L\tau_{op}},\frac{\hbar}{\varepsilon_M\tau_{op}},\frac{\hbar}{U\tau_{op}}\ll1,\label{eq:condition1}
\end{equation}
where $\tau_{op}$ is the operational time to take a braid step.

\textit{Conclusions.---} We have studied the low-energy spectrum of a hybrid system consisting of a topological nanowire hosting a pair of MZM at its opposite ends, and a QD simultaneously coupled to both them. It was demonstrated that the Coulomb interaction between the dot and the superconducting section leads to an additional hybridization of the MZMs, which strongly affects the density of states of the system and modifies the corresponding conductance profiles. It is shown that the interactions compromise the analysis of the topological quality factor, introduced as a quantitative measure of the level of the degree of topological protection in noninteracting system. This leads to the additional constraint for braiding operation in quantum computing process, which should be performed diabatically with respect to the Coulomb charging energy.

\begin{acknowledgments}
 \textit{Acknowledgements.---} LSR acknowledges support from S\~ao Paulo Research Foundation (FAPESP), grant 2015/23539-8. JES acknowledges support from the Coordenação de Aperfeiçoamento de Pessoal de Nível Superior
- Brasil (CAPES) - Finance Code 001 (Ph.D. fellowship). YM and IAS acknowledge support from the Ministry of Science and Higher Education of Russian Federation, goszadanie no. 2019-1246, and ITMO 5-100 Program. ACS acknowledges support from Brazilian National Council for Scientific and Technological Development (CNPq), grant~305668/2018-8.
\end{acknowledgments}

\providecommand{\noopsort}[1]{}\providecommand{\singleletter}[1]{#1}%

\end{document}